\documentclass{ecai}
\usepackage{times}
\usepackage{graphicx}
\usepackage{latexsym}

\usepackage[utf8]{inputenc} 
\usepackage[T1]{fontenc}    
\usepackage{hyperref}       
\usepackage{url}            
\usepackage{booktabs}       
\usepackage{amsfonts}       
\usepackage{nicefrac}       
\usepackage{microtype}      
\usepackage{lipsum}
\usepackage{multicol}
\usepackage{rotfloat}
\usepackage[linesnumbered,ruled,lined,boxed]{algorithm2e}
\usepackage{caption}
\usepackage{subcaption}
\usepackage{amsmath}

\usepackage{fancyhdr}

\newcommand{\eightrm}{}
\begin{document}

\title{Tree search algorithms for the Sequential Ordering Problem}

\author{Luc Libralesso \and Abdel-Malik Bouhassoun \and Hadrien Cambazard \and
Vincent Jost%
\institute{Univ. Grenoble Alpes, CNRS, Grenoble INP, G-SCOP, 38000 Grenoble, France, email: luc.libralesso@grenoble-inp.fr}
}
%
%
%
\maketitle              
\begin{abstract}
We present a study of several generic tree search techniques applied to the \emph{Sequential Ordering Problem}.
This study enables us to propose a simple and competitive tree search algorithm. It consists of an iterative Beam Search algorithm that favors search over inference and integrates dynamic programming inspired cuts. It proves optimality on half of the SOPLIB instances and finds new best known solutions on 6 among 7 open instances of the benchmark in a small amount of time.

\end{abstract}

\section{INTRODUCTION}\label{sec:intro}

\paragraph{}
Many optimization algorithms rely on ``Branch and bound'' techniques. In most applications, \emph{Depth First Search} or \emph{Best First Search} are used as a default search strategies. \cite{t2004revisiting} shows that in the scheduling context, the choice of these strategies is strongly questionable. In this paper, we draw similar conclusions on the Sequential Ordering Problem.

\paragraph{}
On branch and bound algorithms specific to the Sequential Ordering Problem, most effort have been done on bounds, cuts and hybridisation with local search based strategies (see \cite{gouveia2015load}, \cite{shobaki2015exact}, \cite{jamal2017solving}). To the best of our knowledge however, the impact of using different search strategies has never been studied in details.

\paragraph{}This paper aims to fill this research gap and studies the search strategy impact on SOP branch and bound algorithms. We discuss and study branch and bound generic building blocks and the performance of each combination. We show that simply choosing the right combination of elementary ideas (mostly coming from \cite{shobaki2015exact}) makes a huge impact and lead to a competitive (and state of the art) method.
Indeed, to drastically improve on \cite{shobaki2015exact} results, it ``suffices'' to replace the Depth First Search by a Beam Search and to disable the Minimum Spanning Tree lower bound.
It turns out that the most efficient combination relies on simple lower bounds maintained in constant time, and very efficient tree explorations using a Beam Search strategy. 
This method is able to find better solutions than the currently best known ones on 6 among 7 open instances of the SOPLIB in a short amount of time (less than 600 seconds on a laptop computer) and outperforms the existing local search and branch and bound based approaches for the Sequential Ordering Problem.
The source code and solutions can be downloaded at

\url{https://gitlab.com/librallu/cats-ts-sop}.

\paragraph{}This paper is structured as follows: Section \ref{sec:intro} presents the Sequential Ordering Problem and a quick survey of existing methods (exact methods and meta-heuristics). Section \ref{sec:bs} presents the SOP specific bounds we use (namely prefix bound, ingoing/outgoing bound, MST bound). Section \ref{sec:ts} presents the generic branch and bounds parts we use (namely DFS, LDS, Beam Search and Prefix Equivalence). Finally, Section \ref{sec:res} presents numerical results on the impact of the search strategy and a comparison with existing state of the art algorithms.

\subsection{SOP formal definition}

\emph{Sequential Ordering Problem (SOP)} is an Asymmetrical Traveling Salesman Problem with precedence constraints.

\paragraph{}An instance of SOP consists of a directed graph $G=(V,A)$, arc weights $w: A \rightarrow \mathbb{R}$, a set of precedence constraints $C \subseteq V \times V$ modeled as another graph, a start vertex $s\in V$, and a destination vertex $t\in V$.

We search for a permutation of vertices that starts with $s$, ends with $t$, satisfies the precedence constraints (\emph{i.e.} for each precedence constraint $(a,b) \in C$, vertex $a$ must be visited before vertex $b$) and that minimizes the weighted sum of the arcs joining the vertices in the permutation.

\begin{figure}[ht]
    \centering
    \begin{subfigure}[b]{0.25\textwidth}
        \centering
        \includegraphics[width=4cm]{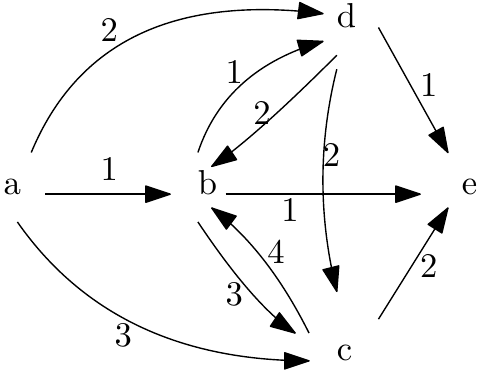}
        \caption{graph representation\\arc weights}
        \label{fig:sop_ex_arcs}
    \end{subfigure}
    \begin{subfigure}[b]{0.25\textwidth}
        \centering
        \includegraphics[width=4cm]{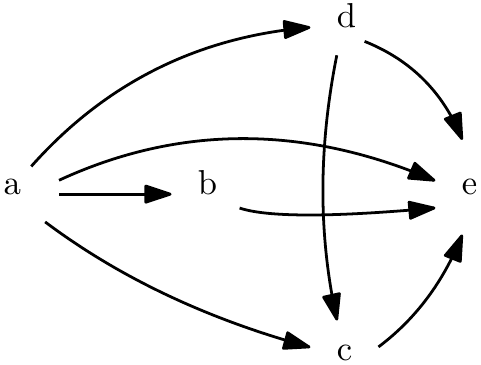}
        \caption{graph representation\\precedence constraints}
        \label{fig:sop_ex_pred}
    \end{subfigure}
    \caption{Example of a SOP instance with 5 vertices and 1 precedence constraint where $a$ is the start vertex and $e$ the end vertex. Permutation $a,d,c,b,e$ is a feasible (since $d$ is visited before $c$) and has cost $2+2+4+2 = 10$.  Permutations $a,b,c,d,e$ and $a,c,b,d,e$ are not feasible. 
    Permutation $a,b,d,c,e$ is optimal with cost $1+1+2+2 = 6$. 
    }
    \label{fig:sop_ex}
\end{figure}

See Figure~\ref{fig:sop_ex} for an illustrative example.


\subsection{Literature review}

\paragraph{}SOP was originally presented in \cite{escudero1988inexact} alongside some exact algorithms based on a mathematical programming model. It has been extensively studied in the past 30 years, and many applications and resolution methods have been considered.
SOP generalizes several combinatorial problems: Relaxing the precedence constraints gives the Asymmetric Traveling Salesman Problem (ATSP) \cite{laporte1995routing}. If arc length are symmetric, we get the symmetrical TSP. We present in this section the most common applications and algorithms for SOP.

\paragraph{}SOP arises in many industrial applications. We briefly present some of them. \cite{ascheuer1996hamiltonian} presents a stacker crane trajectory optimization. It has to fulfill transportation jobs as fast as possible. This problem can be modeled using SOP where vertices represent jobs and arcs weights represent the time needed to go from a job to another. \cite{spieckermann2004sequential} presents an application of SOP in automotive paint shops where the goal is to minimize the set-up cost of a paint job (flushing old paint, retrieving new color \emph{etc.}). Also, since car lanes relative order cannot be changed during retrieval, precedence constraints need to be taken into account. 
\cite{shobaki2015exact} shows that SOP occurs in the switching energy minimization of compilers. While compiling a program, the compiler has to visit operations so that the switching cost is minimized. Since some operations require other operations to be done before starting, precedence constraints also need to be considered. One can also note the use of SOP in freight transportation \cite{escudero1994lagrangian}, flexible manufacturing systems \cite{ascheuer1996hamiltonian}, and helicopter visiting \cite{fiala1992precedence}.

\paragraph{}Many exact approaches have been proposed to solve the Sequential Ordering Problem. As we discuss in this section, most of the literature focuses on finding strong lower bounds. Earlier approaches to SOP include cutting planes \cite{ascheuer1993cutting},  Lagrangian relax and cut algorithm \cite{escudero1994lagrangian}. A mathematical programming model solved with a branch and bound in which the branching is performed in order to decompose the problem as much as possible was also studied \cite{mojana2012branch}.
The uncapacitated m-PDTSP, which is a generalization of SOP, led to competitive results on SOP using a branch and cut algorithm combined with a generalized variable neighborhood search \cite{gouveia2015load}. Also, decision diagrams made a huge impact by generating automatically good quality bounds  \cite{cire2013multivalued,hernadvolgyi2004solving}. In 2015, a dedicated branch and bound has been proposed \cite{shobaki2015exact}, it combines quick and elementary bounds (prefix, ingoing/outgoing degrees and MST) with a technique inspired from TSP dynamic programming called \emph{History Cuts} that allows to cut dominated partial solutions. Despite the simplicity of its bounds, the later method obtained excellent numerical results. It therefore inspired us to study further the impact of the branch and bound components. This algorithm has been further improved in \cite{jamal2017solving} by the integration of a custom assignment bound and a local-search at each node of the search tree.

\paragraph{}In meta-heuristics, numerous works focus on a local search move called SOP-3-exchange and combine it with various searches. It is a 3-OPT move optimized to take into account precedence constraints and asymmetrical arc weights. This SOP-3-exchange procedure is presented in~\cite{gambardella1997has} alongside an Ant Colony Optimization algorithm. It has also been used within a particle swarm optimization algorithm~\cite{anghinolfi2011hybrid}, by a hybrid genetic algorithm using a new crossover operator referred to as~\emph{Voronoi Quantized Crossover}~\cite{seo2003hybrid}, as well as a bee colony optimization~\cite{wun2014bee}, and a parallel roll-out algorithm~\cite{guerriero2003cooperative}.

Since the hybrid ant colony algorithm HAS-SOP proposed in \cite{gambardella1997has} obtained excellent numerical results, a considerable amount of work have been done to improve it. \cite{gambardella2012enhanced} improves it by the integration of a better data structure called the \emph{don't push stack}. HAS-SOP was again improved in \cite{skinderowicz2017improved} by the integration of a Simulated Annealing scheme. Recently, \cite{helsgaun2017extension} improved the LKH heuristic and made an extension to solve SOP instances. These two last methods obtained the best solutions on large instances of the SOPLIB.

\paragraph{}According to the literature review on the Sequential Ordering Problem, the existing works seem to consider as a working hypothesis that local search is indispensable to obtain state of the art solutions on large instances and that strong lower bounds are the key components of Branch and Bound algorithms. In the next sections of this paper, we investigate different branch and bound components and show that specific combinations can build very efficient methods that are competitive with state of the art meta-heuristics and provide new best known solutions on large SOPLIB instances. Moreover, this heuristic method is also able to prove optimality if the instance is highly constrained, which is not possible with most local search strategies.

\section{AN IMPLICIT SEARCH TREE FOR SOP}\label{sec:bs}

\paragraph{}When designing a tree search algorithm, it is common to divide it into two parts. The implicit search tree (problem specific part, \emph{i.e.} how to branch, bounds, cut, \emph{etc.}) and the generic parts (a search strategy, such as DFS, Beam Search \emph{etc.} or generic cuts, in our case domination cuts). This section presents the problem specific parts and the next section presents the generic parts.

\paragraph{}During the implementation of the implicit search trees, we focused on fast bounds ($O(1)$ for ingoing/outgoing bounds and $O(|E|)$ for the Minimum Spanning Tree bound). The key idea is to favor search over bounding/filtering. We show that quick bounds enable us to design a branch and bound that is competitive with state of the art meta-heuristics. In the specific case of SOP, we show that using stronger bounds dramatically affects the performance of the method, even if the resulting branch and bound explores a smaller tree and has a better guidance.

\subsection{Branching Scheme}
We branch as follows: The root node contains the start vertex $s$. Each child of a given node corresponds to each possible next vertex to be visited (vertices not already added to the prefix and whose predecessors have all been added already to the prefix).

\subsection{Definition and computation of lower bounds}
We define our bounds as it is usually done in AI Planning. For a given node $n$ we define the lower bound $f(n)$  as follows:

\[ f(n) = g(n) + h(n) \]

where:
\begin{itemize}
    \item $g(n)$ is the prefix bound (\emph{i.e.} cost of arcs between already selected vertices)
    \item $h(n)$ is the suffix bound (\emph{i.e.} an optimistic estimate of the remaining work to be done). The three bounds we develop in this section only differ on this criterion.
\end{itemize}

\subsection{Prefix bound}

The prefix bound consists in setting $f(n)=g(n)$ for any node of the search tree. That is $h(n)=0$.

This bound (\emph{i.e.} $g(n)$) can be computed in $O(1)$ along a branch of the search tree, simply by accessing, when adding vertex $b$ to a prefix that ended with vertex $a$, the cost $w_{ab}$ from the input. Within the scope of validity of our computational experiments, and despite its simplicity, this bound revealed itself as the best among the three bounds considered.

\subsection{Ingoing/Outgoing bound}

For the Ingoing/Outgoing (or I/O) bound, we keep the prefix bound and add a lower bound on the suffix.

Consider a node $n$ of the search tree. Let $\text{prefix}(n) = v_1 \dots v_{k-1}$ be the an ordered set of already visited vertices excluding the last added vertex $v_k$, and $\text{suffix}(n)$ the set of remaining vertices to add. We remind that $s$ denotes the start vertex and $t$ the end vertex of the SOP instance.

We design the optimistic estimate of remaining cost $h(n)$ as follows:

\[ h(n) = \max( h_{in}(n), h_{out}(n) ) \]

where:

\[ h_{in}(n) = \sum_{v \in \text{suffix(n)}} \min_{u \in V, uv \in A} w_{uv} \]

\[ h_{out}(n) = \sum_{u \in \text{suffix(n)} \cup \{v_k\} \setminus \{t\} } \min_{v \in V, uv \in A} w_{uv} \]

\paragraph{}
This bound can be computed in $O(1)$ along a branch of the search tree. Indeed, one can precompute the sum of ingoing arcs at root node. When adding a vertex $v$ to the prefix, this sum can be updated in constant time by removing the minimum ingoing arc for $v$. 
The same algorithm can be applied for outgoing arcs. 

\subsection{Minimum spanning tree bound} 
For the MST bound, we keep the prefix bound and add a lower bound on the suffix.

Let $w_{ab}=+\infty$ if $b$ must be visited before $a$. Define $w'_{ab} = \text{min}(w_{ab}, w_{ba})$. The suffix cost $h(n)$ is then computed using Prim's algorithm on the graph spanned by the vertices not yet visited, with edge costs $w'$. Since SOP instances are defined by very dense graphs, there is no complexity difference between our implementation and a variant of Kruskal's algorithm. 

A key analysis, on the Instances used for this paper, revealed that it would be pointless to try to speed-up the implementation of the MST bound, because, even it could be computed as fast as the prefix bound, it would not lead to better solutions than the  algorithms using this weaker bound. 
To prove this, run an algorithm with MST within the time limit. Then run algorithms with cheaper bounds restricting the number of nodes to the number opened by the MST based algorithm.



\section{TREE SEARCH GENERIC COMPONENTS}\label{sec:ts}

In this paper, we examine several tree search components (namely the search strategy and the prefix equivalence, which can be seen as a form of dynamic programming or no-good recording embedded within a branch and bound scheme). We present a comprehensive study of the impact of these blocks and provide an efficient method based on this study.


\subsection{Search strategies}

Consider an implicit search tree which is composed of a root node, bounds, and a children generation procedure for each node. The search strategy explores this implicit tree and aims to find the best possible solution and explore the whole tree which implies proving optimality. Since the 60s, new and efficient search strategies have been published. In this section, we describe some popular strategies within tree search algorithms that we use in our analysis.

\subsubsection{Depth First Search}
(DFS) explores a tree starting by the most promising child, explores the corresponding sub-tree entirely and eventually goes to the next child. This algorithm consumes a limited amount of memory while running ($O(nd)$ where $n$ is the maximum number of children per node and $d$ is the maximal depth of the tree). 

However, DFS suffers from bad decisions made early in the tree exploration. Indeed, the search trees are usually so large that it is virtually impossible for DFS to overcome a bad decision taken at the root node. We note that many mechanisms such as random restarts or search strategies such as LDS have been designed to compensate this drawback of DFS. We precisely focus in this paper on such search strategies.

\subsubsection{Limited Discrepancy Search}
(LDS) was originally proposed in \cite{harvey1995limited}. Given a maximum number of allowed discrepancies $d$, an iteration of LDS explores all nodes that have at most $d$ deviations from the best child according to the guide (in our case the lower bound $f(n)$). Each node stores an allowed number of discrepancies. The root node starts with $d$ allowed discrepancies. Its first best child is given $d$ allowed discrepancies, its second best $d-1$ and so on. Nodes with negative discrepancies are not considered and cut. This allows to explore the most promising branches of the tree while performing a restricted exploration of the other branches. It usually gives better solutions than DFS but can miss the optimal solutions. If $d=1$, LDS behaves like a greedy algorithm. If $d = \infty$, LDS behaves like DFS.

Algorithm \ref{alg:lds} shows the pseudo-code of an iterative LDS algorithm. We start by 1 allowed discrepancy. When the search ends, we restart with 2 allowed discrepancies until the stopping criterion is met.

        \begin{algorithm}[ht]
            \SetAlgoLined
			\SetKwInOut{Input}{Input}
			\SetKwInOut{Output}{Output}
			\Input{$G=(V,A)$, precedence constraints}
			\Output{permutation of $V$}
			\BlankLine
			$d \gets $ 1\;
			\While{stopping criterion not met}{
			    $root.d \gets d$\; 
				Stack $\gets$ root\;
                \While{Stack $\neq \emptyset$}{
                    $n \gets $ Stack.pop()\;
                    $i \gets 0$\;
                    \For{$c \in sortedChildren(n)$}{
                        $c.d \gets n.d - i$\;
                        Stack.push($c$)\;
                        \If{$c.d = 0$}{
                            break\;
                        }
                        $i \gets i+1$\;
                    }
                }
                $d \gets d + 1$\;
			}
            Report best solution found\;
        \caption{Iterative LDS algorithm \label{alg:lds}}
        \end{algorithm}

\subsubsection{Beam Search}
In LDS, nodes are selected depending on a comparison with their siblings and not depending on their absolute quality. We now present \emph{Beam Search (BS)} that aims to explore a subset of a tree that only keeps the best nodes at a given level.
Beam Search has been used successfully to solve many scheduling problems \cite{ow1988filtered,sabuncuoglu1999job}.
Beam Search is a tree search algorithm that uses a parameter called the beam size ($D$). Beam Search behaves like a truncated \emph{Breadth First Search (BFS)}. It only considers the best $D$ nodes on a given level. The others are discarded. Usually, we use the bound of a node to choose the most promising nodes. It generalizes both a greedy algorithm (if $D=1$) and a BFS (if $D=\infty$).
   
Algorithm \ref{alg:BS} shows the pseudo-code of an iterative beam search. The algorithm runs multiple beam searches starting with $D=1$ (line 1) and increases geometrically the beam size (line 8). Each run explores the tree with the given parameter $D$. At the end of the time limit, we report the best solution found so far (line 10).
        
        \begin{algorithm}[]
            \SetAlgoLined
			\SetKwInOut{Input}{Input}
			\SetKwInOut{Output}{Output}
			\Input{$G=(V,A)$, precedence constraints}
			\Output{permutation of $V$}
			\BlankLine
			$D \gets $ 1\;
			\While{stopping criterion not met}{
				Candidates $\gets$ root\;
                \While{Candidates $\neq \emptyset$}{
                    Children $\gets \{ \;\; children(n) \;\;|\;\; n \in \;\;$Candidates$ \;\; \}$ \;
                    Candidates $\gets$ best $D$ nodes among Children\;
                }
                $D \gets D \times $ 2\;
			}
            Report best solution found\;
        \caption{Iterative Beam Search algorithm \label{alg:BS}}
        \end{algorithm}

\subsection{Prefix equivalence cuts}

\paragraph{}Prefix equivalence cuts are a way to eliminate symmetries and dominated partial-solutions. It can be seen as a form of dynamic programming integrated within a tree search algorithm. It stores all (in some variants only a subset) of explored sub-states. Each node compares its prefix subset and last vertex to existing entries in the database. If it is dominated, the node is cut. This strategy has been used in a large variety of methods. For instance, memorisation in branch and bounds (\cite{t2004revisiting,shobaki2015exact,shang2018memorization}), as a form of no-good recording in Constraint Programming (\cite{schiex1994nogood}) or pattern databases (\cite{culberson1998pattern}).

\paragraph{}A prefix equivalence cut for the Sequential Ordering Problem can be defined as follows (inspired from the TSP dynamic programming from \cite{chauhan2012survey}, history cuts from \cite{shobaki2015exact} and the call-based dynamic programming from \cite{benoist2014call}):

Two solution prefixes $n_1,n_2$ are called \emph{equivalent} if they cover the same subset $S \subseteq V$ of vertices and end with the same last vertex $v$. If the prefix cost $g(n_1)$ (\emph{i.e.} the sum of selected arcs between vertices from $S \cup \{v\}$) is (strictly) greater than $g(n_2)$, then $n_1$ is (strictly) dominated by $n_2$ and thus can be cut.

\paragraph{}In other words, the Prefix Equivalence cuts can be seen as a form of dynamic programming where the formulation can be described as follows where pred($S,j$) indicates that $j$ is not a predecessor of any vertex in $S$:

\[ f^*(S,i) = \text{min}_{j \in S \wedge \text{pred}(S,j)} ( f^*(S \setminus \{j\}, j) + c_{ji} \]

\paragraph{}Our implementation of Prefix equivalence consists in altering the behaviour of the branch and bound as follows: Each time a node $n$ is opened, the prefix of $n$ is compared to what exists in the database. If the subset of vertices spanned by $n$  does not exist in the database it is added to it, otherwise it is compared to the best equivalent prefix found so far. The best of the two is kept in the database and the other is discarded.

We implement the database using a hash table. In our numeric experiments, we notice that a branch and bound using the prefix equivalence opens in average 4 to 5 times less nodes than its equivalent version without prefix equivalence.

\paragraph{}
In some versions of the Prefix Equivalence (for instance the one found in \cite{shobaki2015exact}), nodes are cut if their prefix matches an existing entry in the database even if their cost is equal. Notice that we perform restarting tree searches (\emph{i.e.} Iterative Beam Search and Limited Discrepancy), which perform heuristic cuts (they prune nodes to avoid saturating the memory and to ensure reaching feasible solutions). To allow our algorithms to close an instance (\emph{i.e.} to prove the optimality of the best solution it found), we prune nodes only if they are \emph{strictly} dominated by the best equivalent recorded in the database. The reason for doing so is that, although the value recorded in the database corresponds to a node that has been already partially explored, this exploration might have been partial and we need to ensure that the search does not perform any heuristic cut when it provides a proof of optimality.






\section{COMPUTATIONAL RESULTS}\label{sec:res}

Results were obtained from a Intel(R) Core(TM) i5-3470 CPU @ 3.20GHz with 8GB RAM. We run each pair of instance-algorithm for 600 seconds. Instances come from the SOPLIB benchmark available here

\url{http://www.idsia.ch/~roberto/SOPLIB06.zip}

The Instances are randomly generated and their names contain $3$ numbers indicating: the number of nodes (from $200$ to $700$), the range of the cost drawn uniformly (either between $0$ and $100$ or between $0$ and $1000$). Notice that the Instance $R.200.100.60$ is ill defined as its cost are drawn between $0$ and $1000$.

Best known bounds and solutions are an aggregation of results coming from \cite{skinderowicz2017improved}, \cite{gouveia2015load}, \cite{mojana2012branch}, \cite{jamal2017solving} and \cite{helsgaun2017extension}.
Note that the Lin Kernighan Helsgaun 3 Algorithm \cite{helsgaun2017extension} was run on each instance for 100.000 seconds.
The Enhanced Ant Colony System with Simulated Annealing \cite{skinderowicz2017improved} was run 30 times per instance for 600 seconds so 18.000 seconds per instance. The time limit of 600s used in the present paper is therefore considerably smaller.

\subsection{Performance of tree search components}

\paragraph{}We ran 18 different tree searches (DFS, LDS and Beam Search) with and without Prefix equivalence using the prefix, Ingoing/Outgoing degrees or the MST bound for 600 seconds. It turns out that there are two clear winners out of these methods (Beam Search + Prefix Equivalence + Prefix or Ingoing/Outgoing bound). Since the results of the two best methods are very similar, we choose to put the emphasis on the simplest one (\emph{i.e.} Beam Search + Prefix Equivalence + Prefix bound). We show in Table \ref{tab:total_perf_comp} that any deviation of search strategy, cuts or bounds lead to a performance drop (except for the Beam Search + Prefix Equivalence + Ingoing/Outgoing degree bound).


\paragraph{Discussion}
As expected, the Prefix Equivalence cuts perform well on many instances and enable to obtain a significant boost of performance on densities 15, 30 and 60. We note that on loosely constrained instances (\emph{i.e.} 1\% of precedence constraints), the prefix equivalence does not help and thus harm performance since tree searches with prefix equivalence open about 4-5 times less nodes per second than their version without it.

The MST based tree searches open less nodes (1.000 to 10.000 times less than the ingoing/outgoing bound). This leads to less solutions found (sometimes none within the time limit) and is overall less efficient. It appears that on medium size instances, the MST bound does not provide a significant guide improvement (and thus harms performance since it is more expensive to compute than the Prefix or Ingoing/Outgoing bound). One might wonder whether a possible incremental evaluation of the MST bound, 
that is, a computation of it along a branch of the search tree taking advantage of the similarity between the MST for a node and the MST for one of its child, would make a difference.
Indeed, the methodological approach that we tried to highlight in this work allowed to answer clearly and strikingly to this question.
For the benchmark we used, such beautiful incremental algorithm would make absolutely no difference. In the best scenario, we would end up with a third algorithm equivalent to our other two champions. We do not report numerical results on this issue here, but restricting the algorithms by the number of nodes, and not by time limit, we observed that the MST bound did not improve the results overall.

We remark that the search strategy also plays an important role while finding good solutions or closing instances within the time limit.
Globally, the Beam Search strategy finds better solutions than LDS which in turns finds better solutions than DFS. Although DFS is able to find the optimal solution and to prove optimality on some instances, it doesn't match the quality of the solutions of either Beam Search or LDS. The main advantage of DFS is that it does not reopen any node, its main drawback is that it struggles to provide good quality solutions fast. In comparison, Beam Search reopens nodes, but by finding very good solutions fast, it is able to prune more nodes and thus, close more instances.
In this study, the beam search strategy (using prefix equivalence) appears to be the best strategy, both for proving optimality and finding the best solutions within the time limit.

\paragraph{}
We compare the \emph{Beam Search + Prefix Equivalence + Prefix or Ingoing/Outgoing bound} against the best solutions reported in the literature by other state of the art algorithms. Our method finds new best known solutions on 6 among 7 open instances of the SOPLIB in a much shorter time than the other algorithms. It also proves optimality quickly on all instances with 30 and 60 percent precedence (about 10 to 100 times faster than the DFS+prefix equivalence+stronger bounds+local search described in \cite{jamal2017solving}). We remark that the proposed method fails to provide good solutions for 1\% precedence due to the poor quality of bounds on these instances that are close to ATSP.

The simplicity and efficiency of our method makes it much more clear what works and what doesn't for SOP, depending on the density of precedence constraints. When evaluating an algorithm using a combination of strategies, in particular the $3$-OPT exchange for SOP, one might overlook the striking effects that are reported in our Table~\ref{tab:total_perf_comp}. From this table, it becomes evident that computing lower bounds related to ATSP (TSP, I/O, MST, branchings, or assignment) is a priori a major mistake for instances with dense precedence constraints. Conversely, for low density instances, building upon oracles related to ATSP seems necessary.

\section{CONCLUSION AND FUTURE WORKS}

\paragraph{}In this paper, we discussed the impact of the search strategy on the performance of a branch and bound. The beam search appeared to be the most efficient method to find near-optimal solutions quickly compared to DFS and LDS. It also proved itself to be the most efficient method to prove optimality and outperformed DFS by proving optimality on 25 instances (DFS proved optimality only on 17 instances) in less than 600 seconds. Out of this analysis, we provide a simple algorithm that outperforms existing algorithms in solving large instances of the SOPLIB and finds new best known solutions on 6 among 7 open instances in a short amount of time. We demonstrated the importance of deconstructing branch and bound algorithms and of the analysis of contribution and computational cost of each separate building block. We also showed that the search strategy should be given more consideration over trying to improve the quality of the bound without assessing the overall behavior of the considered algorithm. Out of this study, it appears that Branch and Bound algorithms have the potential to compete with classical meta-heuristics if the search is favoured over node inference, and if the search strategy is designed to ensure a good anytime behavior. To this end, other search strategies than the classical Depth First Search or Best First Search should be considered. 
Beam search is a good starting and reference point, as it is both very simple, and within the scope of our paper, undefeated by other tree searches.

\paragraph{}Since the methods highlighted in this paper do not compete on low density precedence constraints with the best methods based on solving the ATSP relaxation of SOP it remains an open question whether and how stronger lower bounds for SOP should be used to drive the search, without dramatically slowing the speed at which nodes are opened.    

\paragraph{}A new benchmark closing the gap between 1\% and 15\% instances could help designing and analysing hybridisation of algorithms competitive in either cases.

\paragraph{}This paper only considers the Sequential Ordering Problem. However, a similar decomposition methodology of complicated algorithms into simple building blocks and the assessment of their contributions, computational costs, and ideally synergies, can be applied on other combinatorial optimization problems. For instance, anytime tree searches have been successfully applied on various hard combinatorial optimization problems such as ``simple assembly line balancing problem'' \cite{blum2008beam}, ``Longest Palindromic Common Sub-sequence'' \cite{djukanovic2019anytime}.

\newcommand\ul[1]{\underline{#1}}
\onecolumn
\begin{table}
    \centering
\begin{tabular}{|l|l|l||l|l|l|l|l|l||l|l|l|}
\hline
Instance                     & BKLB    & BKUB    & BS,PE,P         & BS,PE,IO          & BS,PE,MST        & BS,P & DFS,PE,P & LDS,PE,P & T record (s) & T opt (s) \\
\hline
\hline

R.200.100.1              & 61      & 61      & 189              & 189              & 299              & 189	    & 283	            & 192	  & - & - \\
R.200.100.15             & 1.792   & 1.792   & 1.792            & 1.792            & 1.887            & 1.796	& 3.740	            & 2.325	  & 19.5 & -  \\
R.200.100.30             & 4.216   & 4.216   & \textbf{4.216}   & \textbf{4.216}   & \textbf{4.216}   & 4.249	& \textbf{4.216}	& 4.216	  & 0.1 & 0.6  \\
R.200.100.60             & 71.749  & 71.749  & \textbf{71.749}  & \textbf{71.749}  & \textbf{71.749}  & 71.749	& \textbf{71.749}	& 71.749  & 0.0 & 0.0  \\
R.200.1000.1             & 1.404   & 1.404   & 2.554            & 2.554            & 3.398            & 2.554	& 3.448	            & 2.684	  & - & -  \\
R.200.1000.15            & 20.481  & 20.481  & \textbf{20.481}  & \textbf{20.481}  & 20.952           & 20.517	& 34.982	        & 25.592  & 16.3 & 547.7	 \\
R.200.1000.30            & 41.196  & 41.196  & \textbf{41.196}  & \textbf{41.196}  & \textbf{41.196}  & 41.728	& \textbf{41.196}	& 41.196  & 0.1 & 0.4  \\
R.200.1000.60            & 71.556  & 71.556  & \textbf{71.556}  & \textbf{71.556}  & \textbf{71.556}  & 71.556	& \textbf{71.556}	& 71.556  & 0.0 & 0.0  \\
\hline                                                           
R.300.100.1              & 26      & 26      & 214              & 214              & 406              & 204	    & 265	            & 225	  & - & -  \\
R.300.100.15             & 3.152   & 3.152   & 3.152            & 3.152            & 3.458            & 3.201	& 5.355	            & 4.081	  & 178.9 & -  \\
R.300.100.30             & 6.120   & 6.120   & \textbf{6.120}   & \textbf{6.120}   & 6.330            & 6.200	& \textbf{6.120}	& 6.120	  & 2.2 & 7.9  \\
R.300.100.60             & 9.726   & 9.726   & \textbf{9.726}   & \textbf{9.726}   & \textbf{9.726}   & 9.726	& \textbf{9.726}	& 9.726	  & 0.0 & 0.0  \\
R.300.1000.1             & 1.294   & 1.294   & 3.080            & 3.080            & 4.784            & 2.888	& 3.551	            & 3.012	  & - & -  \\
R.300.1000.15            & 29.006  & 29.006  & 29.006           & 29.006           & 33.885           & 29.481	& 51.152	        & 43.597  & 220.0 & -  \\
R.300.1000.30            & 54.147  & 54.147  & \textbf{54.147}  & \textbf{54.147}  & 54.491           & 54.533	& 55.791	        & 54.147  & 0.4 & 3.6  \\
R.300.1000.60            & 109.471 & 109.471 & \textbf{109.471} & \textbf{109.471} & \textbf{109.471} & 109.471	& \textbf{109.471}	& 109.471 & 0.0 & 0.0  \\
\hline                                                           
R.400.100.1              & 13      & 13      & 191              & 191              & -                & 175	    & 295	            & 203	  & - & -  \\
R.400.100.15             & 3.879   & 3.879   & 3.879            & 3.879            & 5.011            & 3.961	& 8.103	            & 6.584	  & 176.7 & -  \\
R.400.100.30             & 8.165   & 8.165   & \textbf{8.165}   & \textbf{8.165}   & 8.210            & 8.183	& \textbf{8.165}	& 8.165	  & 0.4 & 2.1  \\
R.400.100.60             & 15.228  & 15.228  & \textbf{15.228}  & \textbf{15.228}  & \textbf{15.228}  & 15.228	& \textbf{15.228}	& 15.228  & 0.0 & 0.0  \\
R.400.1000.1             & 1.343   & 1.343   & 3.247            & 3.247            & -                & 3.247	& 4.466	            & 3.525	  & - & -  \\
\textbf{R.400.1000.15}   & 35.364  & 38.963  & 38.963           & 38.963           & 53.789           & 39.722	& 76.463	        & 69.342  & 157.2 & -  \\
R.400.1000.30            & 85.128  & 85.128  & \textbf{85.128}  & \textbf{85.128}  & 87.698           & 85.720	& \textbf{85.128}	& 85.128  & 0.5 & 1.8  \\
R.400.1000.60            & 140.816 & 140.816 & \textbf{140.816} & \textbf{140.816} & \textbf{140.816} & 140.816	& \textbf{140.816}	& 140.816 & 0.0 & 0.0  \\
\hline                                                           
R.500.100.1              & 4       & 4       & 267              & 275              & -                & 202	    & 272	            & 232	  & - & -  \\
\textbf{R.500.100.15}    & 4.628   & 5.284   & \ul{5.261}       & \ul{5.261}       & 7.593            & 5.411	& 9.917	            & 9.610	  & 206.5 & -  \\
R.500.100.30             & 9.665   & 9.665   & \textbf{9.665}   & \textbf{9.665}   & 10.388           & 9.778	& 10.999	        & 9.665	  & 1.4 & 6.2  \\
R.500.100.60             & 18.240  & 18.240  & \textbf{18.240}  & \textbf{18.240}  & \textbf{18.240}  & 18.257	& \textbf{18.240}	& 18.240  & 0.0 & 0.1  \\
R.500.1000.1             & 1.316   & 1.316   & 4.079            & 4.079            & -                & 3.541	& 4.703	            & 3.717	  & - & -  \\
\textbf{R.500.1000.15}   & 43.134  & 49.504  & \ul{49.366}      & \ul{49.366}      & 71.888           & 50.624	& 103.985	        & 94.625  & 120.8 & -  \\
R.500.1000.30            & 98.987  & 98.987  & \textbf{98.987}  & \textbf{98.987}  & 115.074          & 99.492	& 114.544	        & 98.987  & 1.7 & 3.8  \\
R.500.1000.60            & 178.212 & 178.212 & \textbf{178.212} & \textbf{178.212} & \textbf{178.212} & 178.355	& \textbf{178.212}	& 178.212 & 0.0 & 0.0  \\
\hline                                                           
R.600.100.1              & 1       &  1      & 289              & 289              & -                & 289	    & 306	            & 246	  & - & -  \\
\textbf{R.600.100.15}    & 4.803   & 5.472   & \ul{5.469}       & \ul{5.469}       & 9.329            & 5.695	& 13.007	        & 10.939  & 160.5 & -  \\
R.600.100.30             & 12.465  & 12.465  & \textbf{12.465}  & \textbf{12.465}  & 12.929           & 12.475	& 13.899	        & 12.465  & 3.1 & 10.3  \\
R.600.100.60             & 23.293  & 23.293  & \textbf{23.293}  & \textbf{23.293}  & \textbf{23.293}  & 23.324	& \textbf{23.293}	& 23.293  & 0.0 & 0.0  \\
R.600.1000.1             & 1.337   & 1.337   & 4.030            & 4.030            & -                & 3.853	& 4.814	            & 4.074	  & - & -  \\
\textbf{R.600.1000.15}   & 47.042  & 55.213  & \ul{54.994}      & \ul{54.994}      & 90.977           & 55.748	& 115.295	        & 108.164 & 183.6 & -  \\
R.600.1000.30            & 126.789 & 126.789 & \textbf{126.798} & \textbf{126.798} & 158.425          & 128.761	& 145.672	        & 126.798 & 1.6 & 7.2  \\
R.600.1000.60            & 214.608 & 214.608 & \textbf{214.608} & \textbf{214.608} & 214.608          & 214.608	& \textbf{214.608}	& 214.608 & 0.1 & 0.1  \\
\hline                                                           
R.700.100.1              & 1       & 1       & 186              & 250              & -                & 186	    & 281	            & 258	  & - & -  \\
\textbf{R.700.100.15}    & 5.946   & 7.021   & \ul{7.020}       & \ul{7.020}       & 11.392           & 7.220	& 13.778	        & 13.206  & 386.9 & -  \\
R.700.100.30             & 14.510  & 14.510  & \textbf{14.510}  & \textbf{14.510}  & 17.125           & 14.632	& 19.655	        & 14.510  & 4.2 & 21.6  \\
R.700.100.60             & 24.102  & 24.102  & \textbf{24.102}  & \textbf{24.102}  & 24.848           & 24.102	& \textbf{24.102}	& 24.102  & 0.2 & 0.5  \\
R.700.1000.1             & 1.231   & 1.231   & 4.403            & 4.403            & -                & 4.042	& 4.629	            & 4.028	  & - & -  \\
\textbf{R.700.1000.15}   & 54.351  & 65.305  & \ul{64.777}      & \ul{64.777}      & 108.627          & 65.775	& 141.544	        & 121.189 & 25.5 & -  \\
R.700.1000.30            & 134.474 & 134.474 & \textbf{134.474} & \textbf{134.474} & 158.327          & 136.073	& 158.613	        & 134.474 & 1.3 & 4.8  \\
R.700.1000.60            & 245.589 & 245.589 & \textbf{245.589} & \textbf{245.589} & 245.688          & 245.752	& \textbf{245.589}	& 245.589 & 0.1 & 0.1  \\
\hline
\hline
nb closed                &         &         & 25               & 25               & 11               & 0 & 17 & 0 & & \\
\hline
\end{tabular}
    \caption{{Tree search components performance.\\
    \\
BKLB (resp. BKUB) refers to the Best Known Lower Bound (resp. Upper Bound) from our literature review.\\
BS,PE,P refers to the combination of Beam Search, Prefix Equivalence and Prefix bound.\\
BS,PE,IO refers to Beam Search, Prefix Equivalence and Ingoing/Outgoing bound.\\
BS,PE,MST refers to the Beam Search with Prefix Equivalence and MST bound.\\
BS,P refers to the Beam Search with the Prefix bound and without Prefix Equivalence.\\
DFS,PE,P refers to Depth First Search with Prefix Equivalence and Prefix bound.\\
LDS,PE,P refers to limited Discrepancy Search with Prefix Equivalence and Prefix bound.\\
``T record'' indicates the time required by BS,PE,P to reach a solution of value BKUB or lower.\\
``T opt'' indicates the time required by BS,PE,P to close the instance.\\
\\
\textbf{Bold instances} are still open.\\
"-" (in column BS,PE,MST) indicate that no solution has been found by the method within 600 seconds.\\
"-" (in columns record (resp. opt)) indicate that no new record (resp. proof of optimality) was found by any of our methods.\\
\textbf{Bold objective values} indicate when the method was able to close the instance within the time limit.\\
\ul{Underlined objective values} indicate an improvement of best known solutions.\\
}}
    \label{tab:total_perf_comp}
\end{table}
\twocolumn

\paragraph{}Moreover, we only considered DFS, LDS and Beam Search as search strategies. Many more exist  (like Beam stack search \cite{zhou2005beam}, BULB \cite{furcy2005limited}, Anytime Focal Search  \cite{cohen2018anytime} \emph{etc.}). Also, one can study other branch and bound components like the recovering step (\emph{i.e.} performing a local search within each node \cite{della2004recovering}) or the probing strategy (starting a greedy algorithm at each node to obtain a better quality estimate).

\paragraph{}To try and evaluate the contribution of such building blocks and ideas on various problems, we started developing a framework for generic Tree-Search, which allows to define a combinatorial problem as a branching scheme, specific cuts and bounds and then to apply generic Tree Search strategies. This might lead to a new standard way (aside Mathematical Programming, Constraint Programming, Local Search \cite{gardi2014mathematical}, Satisfiability of Boolean Clauses, \emph{etc.}) to define and solve combinatorial problems. 
We are working on this issue, trying, first, to match state-of-the-art algorithms for some specific but well studied problems. This paper reports such a preliminary but striking success story, striking because it proved that more care should be given  to make algorithms simpler, rather than ever more  intricate. In this suggests that Occam's razor still has not been digested fully by the (computer) science community.

\paragraph{Acknowledgments:}
We would like to thank Louis Esperet for his useful comments.

\bibliographystyle{ecai}
\bibliography{main}

\end{document}